\documentclass[prc,twocolumn,showpacs,amsmath,amssymb,superscriptaddress,floatfix,nofootinbib]{revtex4}
\usepackage{mathrsfs,bm}
\usepackage{longtable,lscape}
\usepackage{txfonts}
\usepackage{amssymb}
\usepackage{indentfirst}
\usepackage{graphicx,,booktabs}
\usepackage{multirow}
\usepackage{color}
\usepackage{amssymb}

\begin{document}
\title{$\bar{K}$ -induced formation of the $f_2(1270)$ and $f_2^{'}(1525)$ resonances on proton targets}

\author{Yin Huang}
\email{huangy2014@lzu.cn} \affiliation{Research Center for Hadron and CSR Physics, Lanzhou University and Institute
of Modern Physics of CAS, Lanzhou 730000,China}
\affiliation{School of Nuclear Science and Technology,
Lanzhou University, Lanzhou 730000, China}
\affiliation{Institute of Modern Physics,
Chinese Academy of Sciences, Lanzhou 730000, China}
\author{Ju-jun Xie }
\email{xiejujun@impcas.ac.cn}
\affiliation{Research Center for Hadron and CSR Physics, Lanzhou University and Institute
of Modern Physics of CAS, Lanzhou 730000,China}
 \affiliation{Institute of Modern Physics,
Chinese Academy of Sciences, Lanzhou 730000, China}
\author{Jun He}
\affiliation{Research Center for Hadron and CSR Physics, Lanzhou University and Institute
of Modern Physics of CAS, Lanzhou 730000,China}
\affiliation{Institute of Modern Physics,
Chinese Academy of Sciences, Lanzhou 730000, China}
\author{Xurong Chen}
\affiliation{Research Center for Hadron and CSR Physics, Lanzhou University and Institute
of Modern Physics of CAS, Lanzhou 730000,China}
 \affiliation{Institute of Modern Physics, Chinese Academy of Sciences, Lanzhou 730000, China}
\author{Hong-fei Zhang}
  \affiliation{School of Nuclear
Science and Technology, Lanzhou University, Lanzhou 730000, China}\affiliation{Institute of Modern Physics, Chinese Academy of
Sciences, Lanzhou 730000, China}

\date{\today}
\begin{abstract}
We investigate the productiong of $f_2(1270)$ and $f_2^{'}(1525)$
mesons in the $K^{-}p\to\Lambda{}f_2(1270)$,$K^{-}p\to\Lambda{}f^{'}_2(1525)$
and $K^{-}p\to{}K^{+}K^{-}\Lambda$ reactions within an effective
Lagrangian approach. For $K^{-}p\to{}f_2\Lambda$ reaction, by considering the
contributions from the $t$-channel $K^{+}$ exchange and $u$-channel nucleon pole,
we get a fairly good description of the experimental measurements about the total
and differential cross sections.  Based on the studies of the $K^{-}p\to{}f_2\Lambda$
reaction, we investigate $K^{-}p\to{}K^{+}K^{-}\Lambda$ reaction including the
contributions from the $f_2(1270)$ and $f'_2(1525)$ mesons decaying into $K^+K^-$ pair.
The total cross sections and invariant mass distribution of the $K p \to K^+K^- \Lambda$
reaction are predicted.   The results can be tested in future experiments and therefore
offer new clues on the nature of these tensor states.
\end{abstract}


\maketitle
\section{INTRODUCTION}
Since the charmonium-like resonance $XYZ$ and the new baryonic
$P_c$~\cite{Aaij:2015tga} states are observed, a lot of experiments
spring up to study the exotic states from Belle, BaBar, BESIII,
LHCb, CDF, D0 and other collaborations.  And people
believe that the traditional convention, the meson is made up
of quark and antiquark as well as baryon is made
up of three quarks, is broken.  Furthermore, it is not surprising to find
out that many low-lying states, even those long believed to be
conventional $q\bar{q}$ (or $qqq$) states, may have large components
of other nature.  These observation has attracted a lot of attention
from the theoretical side.  Various explanations of these states have
been proposed, such as molecules, muti-quark states,
kinematic effects, or mixtures of components of different
nature. Nevertheless, up to now none of them has
been accepted unanimously.

Indeed, it has been shown that many of the lowlying
mesonic states can be understood not only as $q\bar{q}$ states but
also as meson-meson molecules, dynamically generated in
the so-called unitary approaches.  For example, $f_2(1270),f_2^{'}(1525)$ and $K_2^{*}(1430)$ are
listed in the Particle Data Group~\cite{K.A. Olive} as the lightest tensor
states and correspond to the $2^{++}$ ground-state nonet.
On the other hand, these states have been studied tensor
mesons in the unitary approach based on the hidden-gauge Lagrangians.
It was found that the $f_2(1270)$, $f'(1525)$, and $K^*_2(1430)$ are dynamically
generated from vector-meson-vector-meson interactions~\cite{Geng:2008gx}.
Within this picture,  the branching ratios into pseudoscalar-pseudoscalar and
vector-vector final states are all consistent with data~\cite{Geng:2008gx}.
In Ref.~\cite{Nagahiro:2008um}, the two-photon decay widths of the
$f_0(1370)$ and $f_2(1270)$ have been calculated and found to agree with data.
Furthermore, the ratios of the $J/\psi$ decay rates into a vector meson ($\phi,\omega$,or $K^{*}$)
and one of the tensor states [$f_2(1270),f_2^{'}(1525)$, and $K_2^{*}(1430)$]
have been calculated, and the agreement with data is found
to be quite reasonable~\cite{MartinezTorres:2009uk}.   Following the same approach, in
Ref.~\cite{Geng:2009iw} it is shown that the ratio of the $J/\psi$ decay rates
into $\gamma{}f_2(1270)$ and $\gamma{}f_2^{'}(1525)$ also agrees with data.
The agreement with experimental data turns out to be quite good in general,
providing support to the underlying assumption that these states contain large
meson-meson components.

Assuming that the $f_2(1270)$ resonance is a $\rho-\rho$ molecular state, it
was found in Ref.~\cite{Xie:2014twa} that the differential cross section as
well as the $t$ dependence of the $\gamma p \to p f_2(1270)$ reaction are
in good agreement with the experimental
results~\cite{Battaglieri:2009aa} and provide support for the molecular
picture of the $f_2(1270)$ resonance in the first baryonic reaction.
In a recent work~\cite{Xie:2015isa}, taking the  $f^{'}_{2}(1525)$
resonances  are dynamically generated states from
vector-meson-vector-meson interactions in the $s$-wave with spin $S = 2$,
the $\gamma{}p\to{}f_2^{'}(1525)p$ reaction has been studied.

In addition to the photoproduction processes, the $\bar{K}N$ scattering can also provide
important information on tensor mesons.
Hence, we study the $K p \to \Lambda f_2$ ($f_2 \equiv f_2(1270),~ f'_2(1525)$) reaction
within an effective Lagrangian approach in this work.   The contributions
from the Born terms including $t-$channel $K^{+}$ exchange and $u$-channel nucleon pole are considered.
Furthermore, for the low energy of the $K^{-}p\to{}K^{+}K^{-}\Lambda$
reaction, we pay especially attention on the role of the $f_2(1270)$ and $f_2^{'}((1525)$ mesons.
\begin{figure}[htbp]
\begin{center}
\includegraphics[scale=0.6]{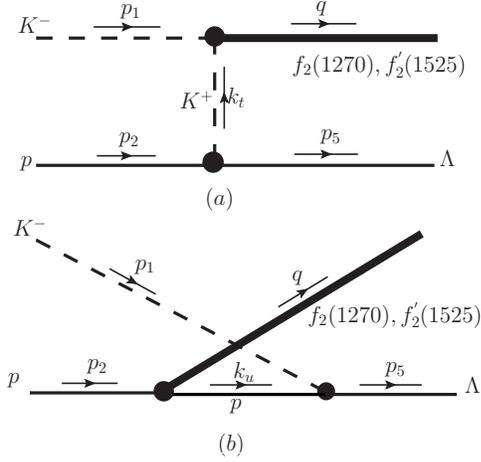}
\end{center}
\caption{Feynman diagrams for $K^{-}p\to{}f_2(1270)\Lambda,f_2^{'}(1525)\Lambda$ reaction.
The contributions from $t$-channel $K^{+}$ exchange (a), $u$-channel nucleon
pole (b).  we also show the definition of the kinematical ($p_1,
p_2, p_3, q,p_5$) that we use in the present calculation. In addition,
we use $k_t=p_1-q$, and $k_u=p_1-p_5$.  }\label{feydiagrams}
\end{figure}
\begin{figure}[h!]
\begin{center}
\includegraphics[scale=0.33]{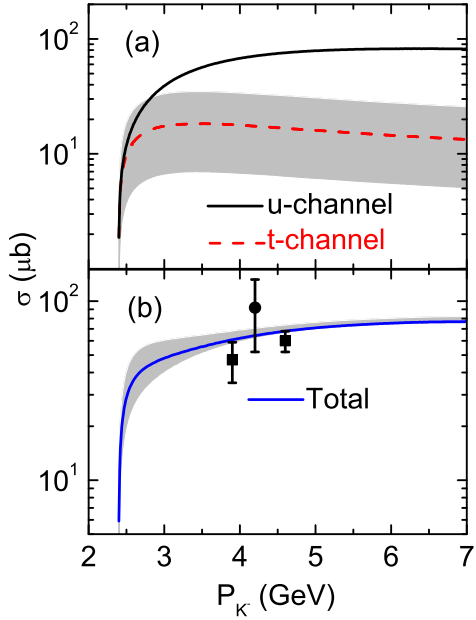}
\end{center}
\caption{(color online). The obtained total cross section for $K^{-}p\to{}f_2(1270)\Lambda$ with the typical cutoff $\Lambda_K=0.51-0.53$ GeV. The experimental data are taken from Ref.~\cite{AguilarBenitez:1980zz}(circle) and Ref.~\cite{AguilarBenitez:1972ey}(square) }\label{cs1}
\end{figure}

\begin{figure}[h!]
\begin{center}
\includegraphics[scale=0.32]{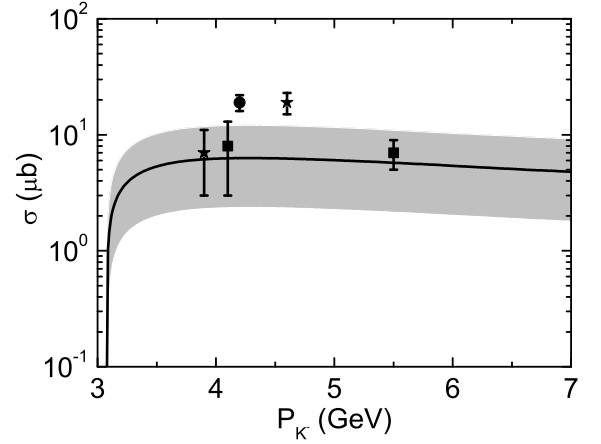}
\end{center}
\caption{The obtained total cross section for $K^{-}p\to{}f_2^{'}(1525)\Lambda$ with the typical cutoff $\Lambda_K=0.51-0.53$ GeV. The experimental data are taken from Ref.~\cite{AguilarBenitez:1980zz}(circle),Ref.~\cite{AguilarBenitez:1972ey}(star) and Ref.~\cite{Mott:1990mb}(square) }\label{cs2}
\end{figure}

This paper is organized as follows.  After the introduction,
we present the calculation of the production of $f_2(1270)$,$f_2^{'}(1525)$
via $\bar{K}$ -induced formation on proton targets.   In Sec. III, the
contribution to the $K^{+}K^{-}\Lambda$ final state is discussed, and the corresponding
total cross sections and invariant mass distribution of these reactions are given. This
work ends with the conclusion.

\section{The $K^{-}p\to{}f_2(1270)\Lambda,f_2^{'}(1525)\Lambda$ reaction}
We choose the production process $K^{-}p$ $\to$ $\Lambda{}f_2$
reactions to study the production of $f_2(1270)$ and $f'_2(1525)$
which can couple to $K^+K^-$~\cite{K.A. Olive}.  First, we mainly
concentrate on the production probability of $f_2$ in the $K^{-}p$
$\to$ $f_2\Lambda$ process, where the
total cross sections of these process are discussed.

The basic tree level Feynman diagrams for the $K^{-}p\to{}f_2\Lambda$ reaction are shown in
Fig.~\ref{feydiagrams}, where the contributions from the $t-$channel $K^{+}$
exchange process [Fig.~\ref{feydiagrams}(a)],  and the nucleon
pole [Fig.~\ref{feydiagrams}(b)] are taken into account.  While the $s$-channel
processes are neglected since the information of those processes is scarce and
we expect these contributions to be small.
\begin{figure}[h!]
\begin{center}
\includegraphics[scale=0.35]{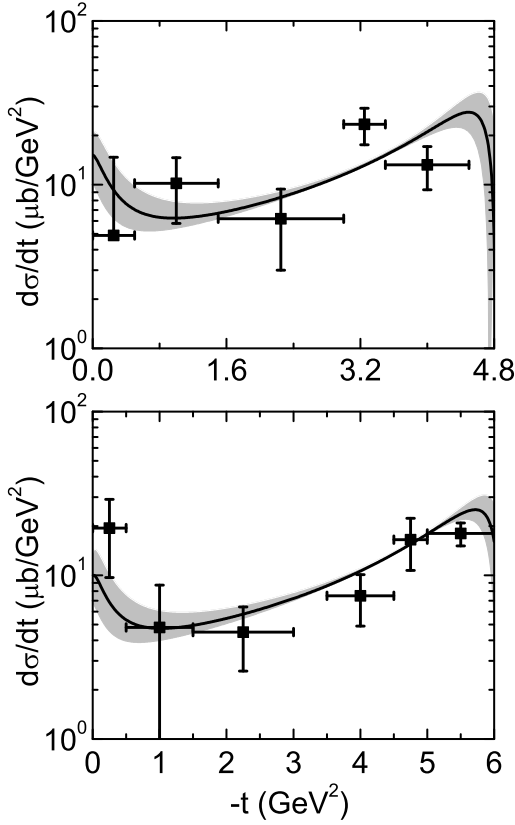}
\end{center}
\caption{The obtained differential cross section for $K^{-}p\to{}f_2(1270)\Lambda$ with the typical cutoff $\Lambda_K=0.51-0.53$ GeV. The experimental data denoted by the squares are taken from Ref.~\cite{AguilarBenitez:1972ey}. }\label{dfs12}
\end{figure}
\begin{figure}[h!]
\begin{center}
\includegraphics[scale=0.31]{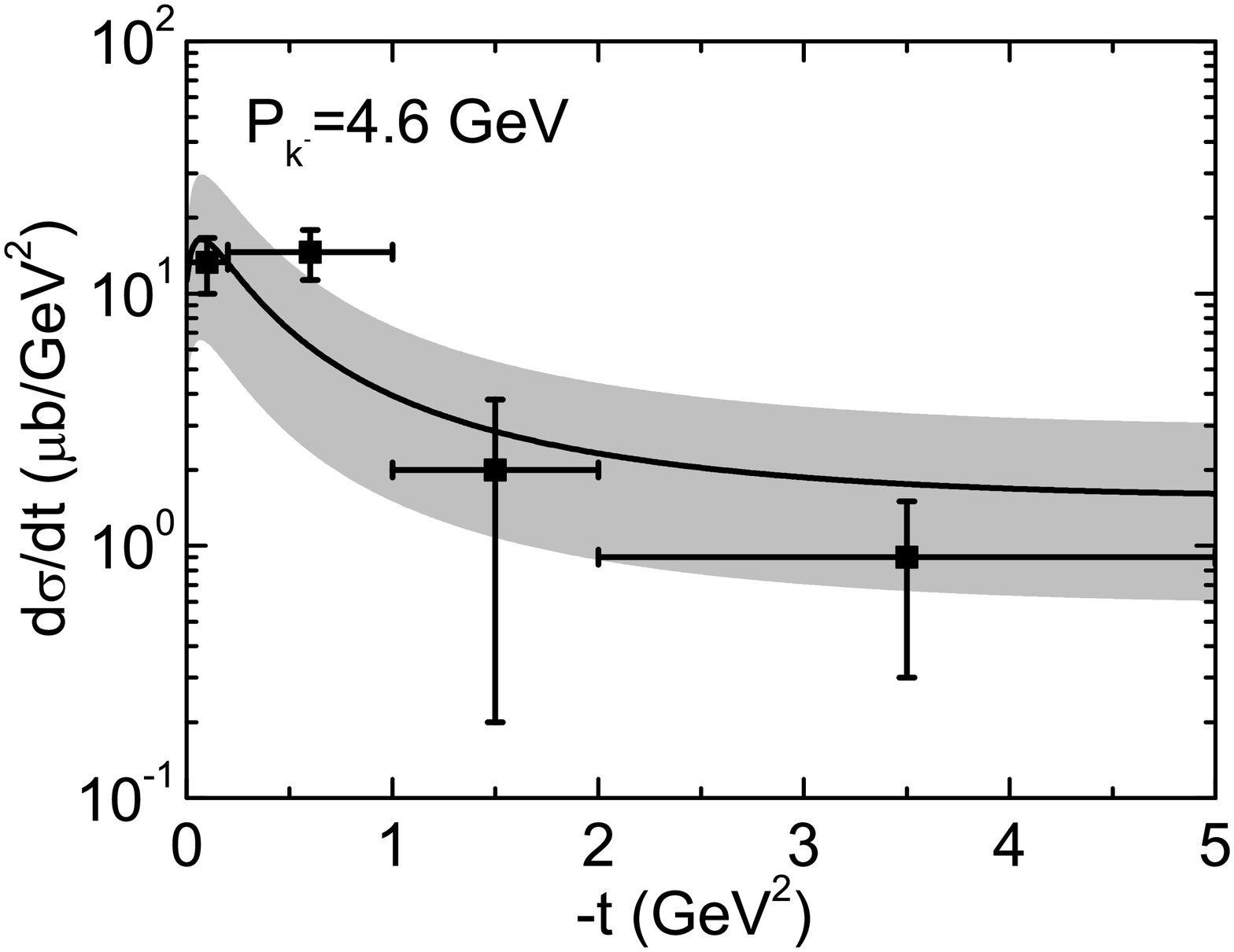}
\end{center}
\caption{The obtained differential cross section for $K^{-}p\to{}f_2^{'}(1525)\Lambda$ with the typical cutoff $\Lambda_K=0.51-0.53$ GeV. The experimental data are taken from Ref.~\cite{AguilarBenitez:1972ey}. }\label{dfs15}
\end{figure}
To compute the amplitudes of these diagrams shown in
Fig.~\ref{feydiagrams}, the effective Lagrangian densities
for the relevant interaction vertexes are needed.
We adopt the effective Lagrangians as used in Refs.~\cite{Oh:2007jd,Yan:1999fn,Goldberg:1968zza,Pilkuhn:1973wq},
\begin{align}
{\cal{L}}_{Kp\Lambda}&=\frac{g_{Kp\Lambda}}{M_N+M_{\Lambda}}\bar{p}\gamma^{\mu}
                     \gamma_{5}\Lambda\partial_{\mu}K^{+}+H.c.,\\
{\cal{L}}_{f_2KK}&=\frac{g_{f_2K^{+}K^{-}}}{2m_{f_2}}f
                     _{2\mu\nu}[K^{-}\partial_{\mu}\partial_{\nu}K^{+}+\partial_{\mu}\partial_{\nu}K^{-}K^{+}\nonumber\\
                     &-2\partial_{\mu}K^{+}\partial_{\nu}K^{-}]+H.c.,\label{Eq1}\nonumber\\
{\cal{L}}_{f_2NN}&=i\frac{G_{f_2NN}}{2M_N}\bar{N}(\gamma_{\mu}\partial_{\nu}+\gamma_{\nu}\partial_{\mu})N
                       f^{\mu\nu}_2+\nonumber\\
                       &+\frac{F_{f_2NN}}{(2M_N)^2}\partial_{\mu}\bar{N}\partial_{\nu}Nf^{\mu\nu}_2+H.c.,
\end{align}
where $M_N$ and $M_{\Lambda}$ are the masses of the nucleon and  $\Lambda(1115)$, respectively.
The $p,\Lambda$,$f_{2\mu\nu}$ are the nucleon field, $\Lambda(1115)$ filed, and
$f_2$ filed, respectively.  The $K$ is kaon field and the isodoublets are defined as
\begin{align}
K&=\left(\begin{array}{cc}
    K^{+}  \\
    K^{0}  \\
  \end{array}\right);
\bar{K}=\left(\begin{array}{c}
    K^{-} ,
   \bar{K}^{0}  \\
  \end{array}\right).
\end{align}
The coupling constant $g_{Kp\Lambda}$ can be determined by flavor
$SU$(3) symmetry relations, which give $g_{Kp\Lambda}=13.24$~\cite{Oh:2007jd}.
While the value of the coupling constant $g_{f_2K^{+}K^{-}}$ can be determined
from the partial decay width of $f_2\to{}K^{+}K^{-}$, which can be obtained from
Eq.~(\ref{Eq1}),
\begin{align}
\Gamma_{f_2\to{}K^{-}K^{+}}&=\frac{m_{f_2}}{480\pi}g^2_{f_2K^{+}K^{-}}(1-\frac{4m_{K}^2}{m_{f_2}^2})^{5/2}\label{Eq2},
\end{align}
where $m_{f_2}$ and $m_{K}$ are mass of $f_2$ and kaon,  respectively.

With mass ($m_{f_2}=1.275$ GeV, $1.525$ GeV, $m_K=0.494$ GeV),
total decay width ($\Gamma_{f_2(1270)}=0.185$ GeV, $\Gamma_{f_2(1525)}=0.073$ GeV),
and decay branching ratio of $f_2\to{}K^{+}K^{-}$ [Br($f_2(1270)\to{}K^{+}K^{-})=0.046$,
Br($f_2(1525)\to{}K^{+}K^{-})=0.887$],  from Eq.~(\ref{Eq2}), we obtain
the coupling constant are shown in Tab.~\ref{Tab:ccs}.

\begin{table}
\caption{The coupling constants $g_{f_2KK}$, $G_{f_2NN}$,and $F_{f_2NN}$ for $f_2\to{}K\bar{K}$ and $N\bar{N}$ reaction.}\label{Tab:ccs}
\begin{tabular}{ccc|ccccccccccccccccccccccccccc}
\hline\hline
&&$Resonance$&  &&&&  &  $channel $ &&&& && $g_{f_2KK}$&   &&&&&  $G_{f_2NN}$&  & && &&     $F_{f_2NN}$ &&  \\\hline
&&$f_2(1270)$&  &&&&  &  $K\bar{K}$&&&&&&   $9.96$&        &&&&&  $--$&         & && &&     $--$        &&   \\
&&$         $&  &&&&  &  $N\bar{N}$&&&&&&   $--$&          &&&&&  $2.19$&       & && &&     $0$         &&  \\
\hline
&&$f^{'}_2(1525)$&&& &&& $K\bar{K}$&&&&&&   $15.78$&       &&&&&  $--$&         & && &&     $--$        &&   \\
&&$             $&&& &&& $N\bar{N}$&&&&&&   $--$&          &&&&&  $ 0$&         & && &&     $0$         &&  \\
\hline \hline
\end{tabular}
\end{table}

By using the tensor meson dominance (TMD)~\cite{Renner:1971mu}, one can determine the universal coupling
constant of the $f_2$ meson from its decay into two pions, which can then be used to determine the coupling
constant for $f_2\to{}N\bar{N}$ reaction (See Ref.~\cite{Oh:2003aw} for details).  And are shown in Table.~\ref{Tab:ccs}.

In evaluating the scattering amplitudes of the $K^{-}p\to{}f_2\Lambda$
reaction, we need to include the form factors because
the hadrons are not pointlike particles. For the
t-channel $K^{+}$ meson exchange, we adopt here a common
scheme used in many previous works~\cite{Kim:2011rm}
\begin{align}
F_{K}(k_t^2)=\frac{\Lambda_K^2-m_k^2}{\Lambda_K^2-k^2_t},
\end{align}
where $\Lambda_{K}$ is the cutoff parameter.

For the $s$-channel and $u$-channel processes, we adopt a
form factor~\cite{Kim:2011rm},
\begin{align}
{\cal{F}}_B(q^2_{ex},M_{ex})=\frac{\Lambda^4_{B}}{\Lambda^4+(q^2_{ex}-M^2_{ex})^2}
\end{align}
where $q_{ex}$ and $M_{ex}$ are the four-momentum and the mass
of the exchanged hadron, respectively.  The cut-off parameter is
constrained between 0.8 and 1.5 GeV for all channels.  For simplicity, we take
$\Lambda_B=0.98$ GeV~\cite{Kim:2011rm} for the $u$-channel nucleon pole,
and the $s$-channel resonance exchanges.

With the above preparation, we finally obtain the amplitude of the
$K^{-}(p_1)p(p_2)\to{}f_2(q)\Lambda(p_5)$ process,
\begin{align}
{\cal{T}}^{a}_{fi}&=\frac{-i}{2m_{f_2}}\frac{g_{Kp\Lambda}g_{f_2K^{+}K^{-}}}{M_N+M_{\Lambda}}\bar{u}(p_5,s_5)k\!\!\!/_t\gamma_{5}u(p_2,s_2)\frac{1}{k_t^2-m_{K}^2}F_{K}(k_t^2)\nonumber\\
                 &\times{}T_{f_2}^{*\mu\nu}(q,\lambda)[-k_{t\mu}k_{t\nu}-p_{1\mu}p_{1\nu}+2k_{t\mu}p_{1\nu}]\label{Eq3}.\\
{\cal{T}}^{b}_{fi}&=-i\frac{g_{Kp\Lambda}G_{f_{2NN}}}{2M_N(M_N+M_{\Lambda})}\bar{u}(p_5,s_5)p\!\!\!/_1\gamma_5\frac{k\!\!\!/_u+M_N}{k_u^2-M_N^2}\nonumber\\
                  &\times(\gamma_{\mu}p_{2\nu}+\gamma_{\nu}p_{2\mu})T^{*\mu\nu}_{f_2}(q,\lambda)u(p_2,s_2){\cal{F}}_N(k_u^2,M_N),
\end{align}
where $s_5,p_5$ and $s_2,p_2$ denote the spin polarization variables and
the four-momenta of the outgoing $\Lambda$ and the initial proton, respectively,
while $q,\lambda$ are the four-momenta and spin polarization variables
of the $f_2$ meson.  The $\bar{u}(p_5,s_5)$ and $u(p_2,s_2)$ are
the Dirac spinors for the $\Lambda$ and proton, respectively, while the
$T_{f_2}^{*\mu\nu}(q,\lambda)$ is the polarization tensor
of the $f_{2}$.

By defining $t=(p_1-q)^2=k_t^2,s=(p_1+p_2)^2$, the
corresponding unpolarized differential cross section
reads as
\begin{align}
\frac{d\sigma}{dt}=\frac{M_NM_{\Lambda}}{16\pi{}s}\frac{1}{|\vec{p}_{1cm}|^2}(\frac{1}{2}\sum_{s_2,s_5,\lambda}|{\cal{T}}_{fi}|^2).
\end{align}
The total cross section can be obtained by integrating over
the range of $|\vec{k}_t|$.

The sum over polarizations, in Eq.~(\ref{Eq2}), can be easily
done thanks to
\begin{align}
\sum_{\lambda}&T_{\mu\nu}(q,\lambda)T_{\alpha\beta}^{*}(q,\lambda)=P_{\mu\nu\alpha\beta}\nonumber\\
              &=\frac{1}{2}(\tilde{g}_{\mu\alpha}\tilde{g}_{\nu\beta}+\tilde{g}_{\mu\beta}
              \tilde{g}_{\nu\alpha})-\frac{1}{3}\tilde{g}_{\mu\nu}\tilde{g}_{\alpha\beta}
\end{align}
for the tensor $f_2$ meson, where $\tilde{g}_{\mu\nu}=-g_{\mu\nu}+\frac{q^{\mu}q^{\nu}}{m^2_{f_2}}$.

As shown in Fig.~\ref{cs1} and~\ref{cs2}, we present the variation of the
total cross sections compare with the experiment data to the different
typical $\Lambda_{K}$ values, where $\Lambda_{K}$ is taken 0.51 GeV to
0.53 GeV for $K^{-}p\to{}f_2(1270)\Lambda$ and $K^{-}p\to{}f_2^{'}(1525)
\Lambda$.  The obtained differential cross section for $K^{-}p\to{}f_2(1270)
\Lambda$ and $K^{-}p\to{}f_2^{'}(1525)\Lambda$ with the typical cutoff
$\Lambda_K$ are also analyzed in Fig~\ref{dfs12} and Fig~\ref{dfs15},
respectively.  The experimental data denoted by the squares are
taken from Ref.~\cite{AguilarBenitez:1972ey}.  We see that our theoretical
results, which is obtained including the contributions from
the $t-$channel $K^{+}$ exchange and nucleon
pole, can give a reasonable description of the experimental data.

\section{The $K^{-}p\to{}f_2(1270)\Lambda,f_2^{'}(1525)\Lambda\to{}K^{+}K^{-}\Lambda$ reaction}
\begin{figure}[h!]
\begin{center}
\includegraphics[scale=0.6]{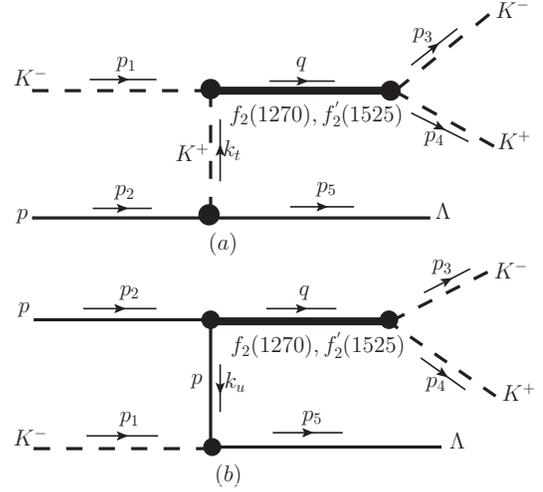}
\end{center}
\caption{Feynman diagrams for $K^{-}p\to{}f_2(1270)\Lambda,f_2^{'}(1525)
\Lambda\to{}K^{+}K^{-}\Lambda$ reaction.  In the diagram, we also show
the definition of the kinematical ($p_1,p_2,p_3,p_4,p_5$) that we use
in the present calculation.}\label{feydiagram23}
\end{figure}
Turning now to the $K^{-}p\to{}f_2\Lambda\to{}K^{+}K^{-}\Lambda$
reaction is described Fig.~\ref{feydiagram23}, the scattering amplitude
${\cal{M}}_{fi}(K^{-}p\to{}K^{+}K^{-}\Lambda)$ is
\begin{align}
{\cal{M}}^{a}_{fi}&(K^{-}p\to{}K^{+}K^{-}\Lambda)=i(\frac{g_{f_2K^{+}
               K^{-}}}{2m_{f_2}})^2\frac{g_{Kp\Lambda}}{M_N+M_{\Lambda}}\nonumber\\
               &\times[p_4^{\mu}(p_3-p_4)^{\nu}-(p_3-p_4)^{\mu}p_3^{\nu}]
               G_{\mu\nu\alpha\beta}^{T}(q)\nonumber\\
               &\times{}[k_{t}^{\alpha}(p_1-k_t)^{\beta}-(p_1-k_t)^{\alpha}p_1^{\beta}]
               \frac{1}{k^2_t-m_{k}^2}\nonumber\\
              &\times\bar{u}(p_5,s_5)k\!\!\!/_t\gamma_{5}
              u(p_2,s_2)F_{K}(k_t^2){\cal{F}}_T(q^2,M_{f_2}),\\
{\cal{M}}^{b}_{fi}&(K^{-}p\to{}K^{+}K^{-}\Lambda)=-i\frac{g_{f_2K^{+}
               K^{-}}}{2m_{f_2}}\frac{g_{Kp\Lambda}}{M_N+M_{\Lambda}}\frac{g_{f_2NN}}{2M_N}\nonumber\\
               &\times[p_4^{\mu}(p_3-p_4)^{\nu}-(p_3-p_4)^{\mu}p_3^{\nu}]
               G_{\mu\nu\alpha\beta}^{T}(q)\nonumber\\
              &\times\bar{u}(p_5,s_5)p\!\!\!/_1\gamma_{5}\frac{(k\!\!\!/_u+M_N)}{k^2_{\mu}-M_N^2}(\gamma_{\alpha}p_{2\beta}+\gamma_{\beta}p_{2\alpha})\nonumber\\
              &\times{}u(p_2,s_2)F_{N}(k_{u}^2){\cal{F}}_T(q^2,M_{f_2}),
\end{align}
where $G_{\mu\nu\alpha\beta}^{T}$ is the propagator of the tensor
meson $f_2$ and can read
\begin{align}
G_{\mu\nu\alpha\beta}^{T}&=\frac{P_{\mu\nu\alpha\beta}}
{q^2-m_{f_2}^2+im_{f_2}\Gamma_{f_2}}.
\end{align}

\begin{figure}[h!]
\begin{center}
\includegraphics[scale=0.32]{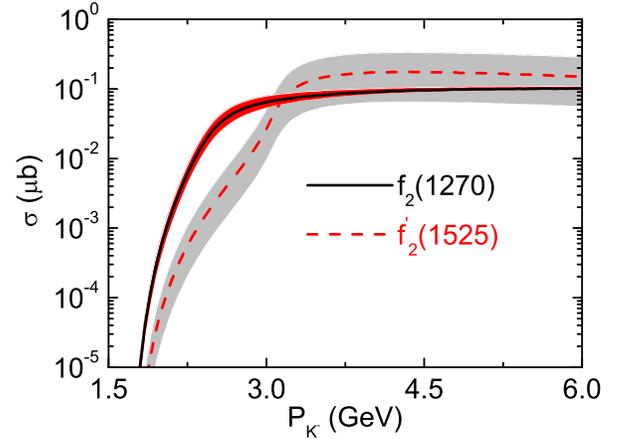}
\end{center}
\caption{(color online). The cross section for $K^{-}p\to{}f_2(1270)
\Lambda,f_2^{'}(1525)\Lambda\to{}K^{+}K^{-}\Lambda$ reaction.}\label{cross23}
\end{figure}

With the formalism and ingredients given above, the cross section versus
 the beam momentum $P_{K^{-}}$ for $K^{-}p\to{}K^{-}K^{+}\Lambda$
reaction is calculated by using a Monte Carlo multiparticle phase
space integration program.  Our predictions, with $\Lambda_K=0.51$ to
0.53 GeV, for the beam momentum $P_{K^{-}}$ from just above
the production threshold 1.7 GeV to 6.0 GeV are shown
in Fig.~\ref{cross23}.  The solid and
dash curves stand for the contributions from the $f_2(1270)$ and
$f_2^{'}(1525)$,  respectively.
\begin{figure*}[h!]
\begin{center}
\includegraphics[scale=0.8]{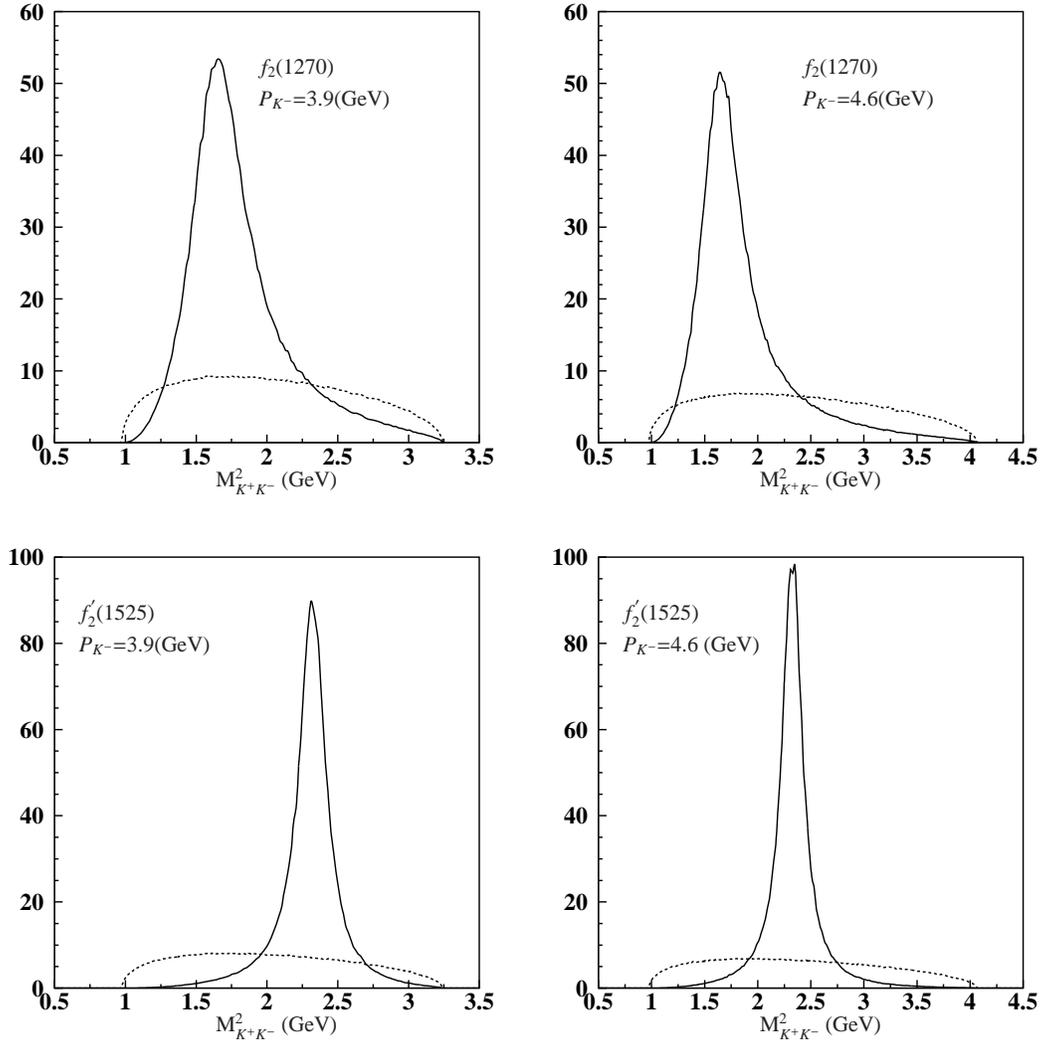}
\end{center}
\caption{The $K^{-}K^{+}$ invariant mass spectrum for the $K^{-}p\to{}K^{+}K^{-}\Lambda$ reaction at
beam momentum $P_{K^{-}}=3.9$ and 4.6 GeV. The dashed lines are pure phase space distributions, while,
the solid lines are full results from our model.}\label{invariantf2}
\end{figure*}

Furthermore, the corresponding $K^{+}K^{-}$ invariant mass spectrum for the $K^{-}p\to{}K^{+}K^{-}\Lambda$
reaction with $\Lambda_K=0.52$ GeV at beam momentum $P_{K^{-}}=3.9$ and 4.6 GeV are calculated and shown in Fig~\ref{invariantf2}.
The dashed lines are pure phase space distributions, while, the solid lines are full results from our
model.   From Fig~\ref{invariantf2}, we can see that there is a clear peak in the $K^{-}K^{+}$ invariant mass
distribution, which is produced by including the contribution from the $f_2(1270)$ and $f^{'}_{2}(1525)$, respectively.

\section{SUMMARY}
In this work, we perform a calculation of the $f_2$[$\equiv f_2(1270)$
$f_2^{'}(1525)$] tenser meson production in the $K^{-}p\to{}f_2\Lambda$ and
$K^{-}p\to{}K^{-}K^{+}\Lambda$ reaction within the effective
Lagrangian method.  For the $K^{-}p\to{}f_2\Lambda$ reaction,
by considering the contributions from the $t$-channel $K^{+}$ exchange and
$u$-channel nucleon pole, we get a fairly good description of the experimental
total cross section data.   Our model shown the
differential cross section $d\sigma/dt$ as function as $-t$,
and  get a good description of the experimental differential cross section data.

Basing on our results of $K^{-}p\to{}f_2\Lambda$ reaction,
we have studied the $K^{-}p\to{}K^{-}K^{+}\Lambda$ reaction.   In this case, we have
considered the contributions from only the $f_2(1270)$ and $f^{'}_{2}(1525)$ mesons.
The invariant mass distribution for the Dalitz
process $K^{-}p\to{}K^{-}K^{+}\Lambda$ shows an obvious peak at $M^2_{K^{+}K^{-}}\approx1.63$
GeV and $M^2_{K^{+}K^{-}}\approx2.33$ GeV,  respectively, which can be checked by further
experiment.

\section*{Acknowledgments}
This project was partially supported by the Major State Basic
Research Development Program in China (No. 2014CB845400), the
National Natural Science Foundation of China (Grants No. 11475227,
No. 11275235, No. 11035006, No.11175220) and the Chinese Academy of
Sciences (the Knowledge Innovation Project under Grant No.
KJCX2-EW-N01, Century program under Grant No. Y101020BR0). It is
also supported by the Open Project Program of State Key Laboratory
of Theoretical Physics, Institute of Theoretical Physics, Chinese
Academy of Sciences, China (No. Y5KF151CJ1).


\begin{thebibliography}{90}
\bibitem{Aaij:2015tga}
  R.~Aaij {\it et al.} [LHCb Collaboration],
  ``Observation of J/¦×p Resonances Consistent with Pentaquark States in ¦«$_b^0$ ¡ú J/¦×K$^-$p Decays,''
  Phys.\ Rev.\ Lett.\  {\bf 115}, 072001 (2015)
  doi:10.1103/PhysRevLett.115.072001
  [arXiv:1507.03414 [hep-ex]].

\bibitem{K.A. Olive}
K.A. Olive et al. (Particle Data Group), Chin. Phys. C, {\bf38},
090001 (2014).


\bibitem{Geng:2008gx}
  L.~S.~Geng and E.~Oset,
  ``Vector meson-vector meson interaction in a hidden gauge unitary approach,''
  Phys.\ Rev.\ D {\bf 79}, 074009 (2009)
  doi:10.1103/PhysRevD.79.074009
  [arXiv:0812.1199 [hep-ph]].

\bibitem{Nagahiro:2008um}
  H.~Nagahiro, J.~Yamagata-Sekihara, E.~Oset, S.~Hirenzaki and R.~Molina,
  ``The gamma gamma decay of the f(0)(1370) and f(2)(1270) resonances in the hidden gauge formalism,''
  Phys.\ Rev.\ D {\bf 79}, 114023 (2009)
  doi:10.1103/PhysRevD.79.114023
  [arXiv:0809.3717 [hep-ph]].

\bibitem{MartinezTorres:2009uk}
  A.~Martinez Torres, L.~S.~Geng, L.~R.~Dai, B.~X.~Sun, E.~Oset and B.~S.~Zou,
  ``Study of the J/psi $\to$ phi(omega) f(2)(1270), J/psi $\to$ phi (omega) f-prime(2)(1525) and J/psi $\to$ K*0(892) anti-K*0(2) (1430) decays,''
  Phys.\ Lett.\ B {\bf 680} (2009) 310
  doi:10.1016/j.physletb.2009.09.003
  [arXiv:0906.2963 [nucl-th]].


\bibitem{Geng:2009iw}
  L.~S.~Geng, F.~K.~Guo, C.~Hanhart, R.~Molina, E.~Oset and B.~S.~Zou,
  ``Study of the f(2)(1270), f(2)-prime(1525), f(0)(1370) and f(0)(1710) in the J/psi radiative decays,''
  Eur.\ Phys.\ J.\ A {\bf 44}, 305 (2010)
  doi:10.1140/epja/i2010-10971-5
  [arXiv:0910.5192 [hep-ph]].

\bibitem{Xie:2014twa}
  J.~J.~Xie and E.~Oset,
  ``Photoproduction of the f$_{2}$(1270) resonance,''
  Eur.\ Phys.\ J.\ A {\bf 51}, 111 (2015)
  doi:10.1140/epja/i2015-15111-3
  [arXiv:1412.3234 [nucl-th]].


\bibitem{Battaglieri:2009aa}
  M.~Battaglieri {\it et al.} [CLAS Collaboration],
  ``Photoproduction of pi+ pi- meson pairs on the proton,''
  Phys.\ Rev.\ D {\bf 80}, 072005 (2009)
  doi:10.1103/PhysRevD.80.072005
  [arXiv:0907.1021 [hep-ex]].

\bibitem{Xie:2015isa}
  J.~J.~Xie, E.~Oset and L.~S.~Geng,
  ``Photoproduction of the $f'_2(1525), a_2(1320)$, and $K^*_2(1430)$,''
  Phys.\ Rev.\ C {\bf 93}, no. 2, 025202 (2016)
  doi:10.1103/PhysRevC.93.025202
  [arXiv:1509.06469 [nucl-th]].

\bibitem{AguilarBenitez:1980zz}
  M.~Aguilar-Benitez {\it et al.} [CERN-College de France-Madrid-Stockholm Collaboration],
  ``Study of Hypercharge Exchange Reactions of the Type 0- 1/2+ $\to$ 2+ 1/2+ at 4-{GeV}/$c$,''
  Z.\ Phys.\ C {\bf 8}, 313 (1981).
  doi:10.1007/BF01546327

\bibitem{AguilarBenitez:1972ey}
  M.~Aguilar-Benitez, S.~U.~Chung, R.~L.~Eisner and N.~P.~Samios,
  ``Study of nonstrange mesons produced in k- p interactions at 3.9 and 4.6 gev/c,''
  Phys.\ Rev.\ D {\bf 6}, 29 (1972).
  doi:10.1103/PhysRevD.6.29


\bibitem{Mott:1990mb}
  J.~Mott {\it et al.},
  ``Study of k-minus p interactions at 4.1 and 5.5 gev/c - final states with two charged particles and a visible lambda,''
  Phys.\ Rev.\  {\bf 177}, 1966 (1969).
  doi:10.1103/PhysRev.177.1966

\bibitem{Oh:2007jd}
  Y.~Oh, C.~M.~Ko and K.~Nakayama,
  ``Nucleon and Delta resonances in K Sigma(1385) photoproduction from nucleons,''
  Phys.\ Rev.\ C {\bf 77}, 045204 (2008)
  doi:10.1103/PhysRevC.77.045204
  [arXiv:0712.4285 [nucl-th]].

\bibitem{Yan:1999fn}
  Y.~Yan and R.~Tegen,
  ``Baryon exchange and meson pole diagrams in p anti-p $\to$ anti-K K, pi+ pi-,''
  Nucl.\ Phys.\ A {\bf 648}, 89 (1999).
  doi:10.1016/S0375-9474(99)00019-6


\bibitem{Goldberg:1968zza}
  H.~Goldberg,
  ``Backward (theta=180-degrees) piN Dispersion Relations: Applications to the Interference Model, P and P-prime Trajectories, and to the Mechanical Form Factors of the Nucleon,''
  Phys.\ Rev.\  {\bf 171}, 1485 (1968).
  doi:10.1103/PhysRev.171.1485

\bibitem{Pilkuhn:1973wq}
  H.~Pilkuhn, W.~Schmidt, A.~D.~Martin, C.~Michael, F.~Steiner, B.~R.~Martin, M.~M.~Nagels and J.~J.~de Swart,
  ``Compilation of coupling constants and low-energy parameters. 1973 edition,''
  Nucl.\ Phys.\ B {\bf 65}, 460 (1973).
  doi:10.1016/0550-3213(73)90296-4

\bibitem{Renner:1971mu}
  B.~Renner,
  ``Empirical test of tensor meson dominance,''
  Phys.\ Lett.\ B {\bf 33}, 599 (1970).
  doi:10.1016/0370-2693(70)90359-X

\bibitem{Oh:2003aw}
  Y.~s.~Oh and T.~S.~H.~Lee,
  ``rho meson photoproduction at low-energies,''
  Phys.\ Rev.\ C {\bf 69}, 025201 (2004)
  doi:10.1103/PhysRevC.69.025201
  [nucl-th/0306033].


\bibitem{Kim:2011rm}
  S.~H.~Kim, S.~i.~Nam, Y.~Oh and H.~C.~Kim,
  ``Contribution of higher nucleon resonances to $K^*{\Lambda}$ photoproduction,''
  Phys.\ Rev.\ D {\bf 84}, 114023 (2011)
  doi:10.1103/PhysRevD.84.114023
  [arXiv:1110.6515 [hep-ph]].




\end{thebibliography}
\end{document}